\documentclass[preprint,12pt]{elsarticle}
\usepackage{graphicx}  
\usepackage{amssymb}
\usepackage{lineno,hyperref}

\journal{Chemical Physics}

\newcommand{\SSCF}{{SSCF}}

\begin{document}
\begin{frontmatter}

\title{Green's functions for spin boson systems: Beyond conventional perturbation theories \tnoteref{VC}}
\tnotetext[VC]{This work is dedicated to Vladimir Chernyak on the occasion of his 60th birthday}
\author[Fd]{Junjie Liu}
\author[Fd]{Hui Xu}
\author[Fd,FF]{Chang-Qin Wu}
\address[Fd]{State Key Laboratory of Surface
Physics and Department of Physics, Fudan University, Shanghai 200433,
China}
\address[FF]{Collaborative Innovation Center of Advanced Microstructures, Fudan University, Shanghai 200433, China}

\date{\today}
\begin{abstract}
Unraveling general properties of Green's functions of quantum dissipative systems is of both experimental relevance and theoretical interest. Here, we study the spin-boson model as a prototype. By utilizing the Majorana-Fermion representation together with the polaron transformation, we establish a theoretical approach to analyze Green's functions of the spin-boson model. In contrast to conventional perturbation theories either in the tunneling energy or in the system-bath coupling strength, the proposed scheme gives reliable results over wide regimes of the coupling strength, bias, as well as temperature. To demonstrate the utility of the approach, we consider the susceptibility as well as the symmetrized spin correlation function (SSCF) which can be expressed in terms of Green's functions. Thorough investigations are made on systems embedded in Ohmic or sub-Ohmic bosonic baths. We found the so-obtained SSCF is the same as that of the non-interacting blip approximation (NIBA) in unbiased systems while it is applicable for a wider range of temperature in the biased systems compared with the NIBA. We also show that a previous perturbation result is recovered as a weak coupling limit of the so-obtained SSCF. Furthermore, by studying the quantum criticality of the susceptibility, we confirm the validity of the quantum-to-classical mapping in the whole sub-Ohmic regime.
\end{abstract}


\end{frontmatter}

\section{Introduction}
In recent years, growing attention has been directed to the realm of quantum dissipative systems, i.e., quantum systems in contact with an environment, owing to the advanced experimental techniques in such distinct fields as condensed matter physics, quantum optics and quantum information. Interpreting and understanding rich phenomena induced by the dissipation from a theoretical point of view are then of fundamental significance. A dissipative two level system(TLS), as a minimal model mimicking the complex dissipative quantum systems, has its own advantages in studying effects of dissipation. On the one hand, due to its diversity of physical realizations, it is capable of describing quantum decoherence in superconducting qubits \cite{Leggett.87.RMP}, tunneling light particles in metals \cite{Fukai.05.NULL}, anomalous low temperature thermal properties in glasses \cite{Anderson.72.PM}, to mention just a few. On the other hand, the system is simple and tractable. Although no exact solutions exist, a number of methods \cite{Egger.92.ZPB,Makri.95.JMP,Bulla.08.RMP,Wang.08.NJP} still produce fruitful results.

A dissipative TLS can be generally described by the spin-boson model(SBM)  \cite{Leggett.87.RMP}, which can be visualized as a spin-$\frac{1}{2}$ degree of freedom with tunneling between the up and down states as well as coupling to a bosonic reservoir built up of harmonic oscillators. In the SBM problem, Green's functions (GFs) or related dynamical correlation functions are quantities of both experimental relevance and theoretical interest. For instance, the symmetrized spin correlation function (SSCF), being proportional to the sum of the greater and lesser GF, can be related to the structure factor \cite{Dattagupta.89.JPCM,Weiss.12.NULL} which determines the inelastic neutron scattering characteristics of the system.

Theoretically, GFs of the SBM have been mostly studied within two main approximation schemes. For small damping strengths, a perturbation expansion in the system-bath coupling strength is appropriate. Due to the absence of an analog of the Wick's theorem for spin operators, standard Feynman diagram
techniques can not be adopted directly. By utilizing the Majorana-Fermion representation (MFR) \cite{Tsvelik.92.PRL}, a diagrammatic method (we refer it to the MFDM in the following) has been put forward and largely developed \cite{Mao.03.PRL,Shnirman.03.PRL,Florens.11.PRB,Schad.15.AP} in this parameter regime. In the opposite regime of moderate to large damping, a perturbation expansion in the tunneling energy applies, yielding the so-called non-interacting blip approximation (NIBA) within the framework of path integrals \cite{Leggett.87.RMP,Weiss.12.NULL}. However, the NIBA breaks down in the presence of a bias at low temperatures. Further extensions are required \cite{Weiss.89.PRL,Goerlich.89.EL,Winterstetter.97.CP,Nesi.07.PRB}. To date, none of theories dealing with GFs can provide a unified description over wide ranges of the parameters space with respect to the system-bath coupling strength, bias and temperature, we have to rely on numerical techniques \cite{Costi.96.PRL,Hofstetter.00.PRL,Bulla.05.PRB,Peters.06.PRB,Anders.07.PRL}. Nevertheless, numerical calculations cannot directly provide intuitive understanding and are often limited to the zero temperature case.

In this work, we propose a unified theoretical scheme, capable of bridging the above two perturbation expansions and meanwhile, removing their shortcomings. Technically, we utilize the MFR such that Wick theorem and standard Feynman diagram techniques can be applied to the SBM. In order to explore the strong system-bath coupling regime, we transform the original Hamiltonian by adopting the polaron transformation (PT) \cite{Weiss.12.NULL} and then perform perturbation expansion. In this sense, our theory can be regarded as an extension to the previous MFDM (thus we refer the proposed scheme to "MFDM+PT"). More importantly, the scheme can be easily extended to study nonequilibrium SBM, especially the heat transport in SBM \cite{Yang.14.EL,Liu.15.NULL}.

Within this scheme, we derive explicit expressions for GFs that are valid ranging from the weak to strong system-bath coupling regime for unbiased or biased systems embedded in Ohmic or sub-Ohmic bosonic baths. We found the so-obtained \SSCF{} is the same as that of the non-interacting blip approximation (NIBA) in unbiased systems while it is applicable for a wider range of temperature in the biased systems. We also show that a previous perturbation result \cite{Shnirman.03.PRL} is recovered as a weak coupling limit of the so-obtained SSCF. Based on the \SSCF{}, we investigate effects of the temperature, bias as well as system-bath coupling strength on the dynamical crossover in the SBM. Using relations between the \SSCF{} and the susceptibility, we further study the excitation spectrum of the spin and the quantum criticality of the susceptibility. We confirm the validity of the critical exponent $\gamma$ obtained for the Ising chain with long-range interactions \cite{Fisher.72.PRL} in the sub-Ohmic SBM, indicating the quantum-to-classical mapping should hold in the system.

The paper is organized as follows. In Sec. \ref{sec:2}, we introduce the SBM and GFs. The dynamical quantities and their connections to GFs are given. We also discuss related issues of the PT. In Sec. \ref{sec:3}, we first briefly review the MFR, as it constitutes the basis for the approach. Then we give details about the calculation of GFs by using the Dyson's equation. In Sec. \ref{sec:4} and \ref{sec:5}, we study the system embedded in the Ohmic and sub-Ohmic bosonic baths, respectively. In Sec. \ref{sec:6}, we summarize our findings and make some final remarks.

\section{Backgrounds}\label{sec:2}
\subsection{Model}
The SBM, consisting of a two level system in contact with a bosonic reservoir, is described by the Hamiltonian \cite{Leggett.87.RMP}
\begin{equation}\label{eq:h}
H~=~\frac{\varepsilon}{2}\sigma_z+\frac{\Delta}{2}\sigma_x+\sigma_z\sum_{j}g_{j}(b^{\dagger}_{j}+b_{j})+\sum_{j}\omega_{j}b^{\dagger}_{j}b_{j},
\end{equation}
where $\varepsilon$ is the energy gap of the two levels, we usually refer it to the bias, $\Delta$ denotes the tunneling energy, $\sigma_{x,z}$ are the Pauli matrices, and $b^{\dagger}_{j}(b_{j})$ denotes the creation (annihilation) operator of the $j$th harmonic mode in the bosonic bath, with $g_{j}$ being the system-bath coupling strength. Throughout the paper, we set $\hbar=1$ and $k_B=1$. The influence of the bath
is contained in the spectral density function
\begin{equation}
J(\omega)=2\pi\sum_{j}g_{j}^2\delta(\omega-\omega_{j}).
\end{equation}
For convenience, we make the specific choice \cite{Weiss.12.NULL}
\begin{equation}\label{eq:bath_spec}
J(\omega)~=~\pi\alpha\omega^s\omega_c^{1-s} e^{-\omega/\omega_c},
\end{equation}
where $\alpha$ is the dimensionless system-bath coupling strength. The case $s>1(s<1)$ corresponds to super-Ohmic (sub-
Ohmic) dissipation, and $s=1$ represents the important
case of frequency-independent (Ohmic) dissipation.

Following the general framework of GFs \cite{Haug.96.NULL}, we introduce the GF for the spin system as
\begin{equation}
\Pi_{\alpha\beta}(t,t^{\prime})~=~-i\left\langle T \sigma_{\alpha}(t)\sigma_{\beta}(t^{\prime})\right\rangle ,
\end{equation}
where $T$ is a time-ordering operator and $\alpha,\beta=x,y,z$. In the equilibrium state, the GF depends only on the time difference, so we introduce $\tau=t-t^{\prime}$. Its retarded, advanced, lesser and greater components are give by
$\Pi_{\alpha\beta}^r(\tau) = -i \Theta(\tau)\langle [\sigma_{\alpha}(\tau),\sigma_{\beta}]\rangle,
\Pi_{\alpha\beta}^a(\tau) = i \Theta(-\tau)\langle [\sigma_{\alpha}(\tau),\sigma_{\beta}]\rangle,
\Pi_{\alpha\beta}^<(\tau) = -i\langle\sigma_{\beta}\sigma_{\alpha}(\tau)\rangle$ and
$\Pi_{\alpha\beta}^>(\tau) = -i\langle\sigma_{\alpha}(\tau)\sigma_{\beta}\rangle$,
respectively, where $\Theta(\tau)$ denotes the Heaviside step function and
$\sigma_{\alpha}(\tau)\equiv e^{iH\tau}\sigma_{\alpha}e^{-iH\tau}$
denotes the operator in the Heisenberg picture.

\subsection{Observables of interest}
For the SBM, one of the dynamical quantities which are of primary interest is the \SSCF{} defined by
\begin{equation}\label{eq:corr_t}
S_z(\tau)~=~\frac{1}{2}\left\langle\sigma_z(\tau)\sigma_z(0)+\sigma_z(0)\sigma_z(\tau)\right\rangle,
\end{equation}
which is directly related to the structure factor \cite{Dattagupta.89.JPCM,Weiss.12.NULL}, thus has an experimental relevance for the inelastic neutron scattering characteristics of the system.
In terms of GFs defined above, the \SSCF{} [Eq. \ref{eq:corr_t}] can be rewritten as a sum of the greater and the lesser GF of $\sigma_z$ in the Fourier space
\begin{equation}
S_z(\omega) ~=~ \frac{i}{2}\left[\Pi_{zz}^>(\omega)+\Pi_{zz}^<(\omega)\right].
\end{equation}

Besides the \SSCF{}, the susceptibility of the spin also has an experimental relevance. The dynamical susceptibility $\tilde{\chi}_z$ (we mark the dynamical susceptibility by a tilde and the static susceptibility by a bar), being the response function of the equilibrated system to an external force coupled to $\sigma_z$, is just the retarded GF of $\sigma_z$
\begin{equation}
\tilde{\chi}_z(\tau)~=~i\Theta(\tau)\langle[\sigma_z(\tau),\sigma_z]\rangle=-\Pi_{zz}^r(\tau).
\end{equation}
Its imaginary part $\tilde{\chi}_z^{\prime\prime}(\omega)$, which determines the energy the spin absorbs from the external field and reflects the excitation spectrum of the spin, can be related to $S_z(\omega)$ via the Fluctuation-dissipation theorem for unbiased systems \cite{Weiss.12.NULL}
\begin{equation}\label{eq:d_sus}
S_z(\omega)~=~\coth\left(\frac{\omega}{2T}\right)\tilde{\chi}_z^{\prime\prime}(\omega).
\end{equation}
It is notable that $\tilde{\chi}_z^{\prime\prime}(\omega)$ is an odd function of $\omega$. Thus the real part $\tilde{\chi}_z^{\prime}(\omega)$ reduces to the static susceptibility $\bar{\chi}_z$ when $\omega=0$, namely, $\bar{\chi}_z=\tilde{\chi}_z^{\prime}(\omega=0)$. Using the Kramers-Kronig relation and the Fluctuation-dissipation theorem, we find at zero temperature that \cite{Bulla.05.PRB,Lu.07.PRB}
\begin{equation}\label{eq:s_sus}
\bar{\chi}_z~=~\frac{2}{\pi}\int_{0}^{\infty}\,d\omega\frac{S_z(\omega)}{\omega}.
\end{equation}
In the vicinity of the quantum critical point $\alpha_c$, the static susceptibility follows \cite{Sachdev.99.NULL}
\begin{equation}\label{eq:s_sus_critical}
\bar{\chi}_z~\propto~|\alpha-\alpha_c|^{-\gamma}
\end{equation}
with $\gamma$ the critical exponent. This critical behavior has attracted many attentions in recent years \cite{Winter.09.PRL,Alvermann.09.PRL,Zhang.10.PRB,Chin.11.PRL,Guo.12.PRL}.

\subsection{Polaron transformation}
Since only GFs of $\sigma_z$ are involved in dynamical quantities we considered , we will mainly focus on the calculation of the GFs of $\sigma_z$ in the following, but the methodologies discussed below can handle other GFs as well with cautions \cite{Schad.16.PRB}.
We principally work here in the so-called scaling limit of $\Delta/\omega_c\ll1$. For such fast baths, it has been demonstrated that the PT can be applied to the entire range of system-bath coupling strength \cite{Lee.12.JCP}. Thus before proceeding to the lengthy calculations, we make the PT with the unitary operator \cite{Weiss.12.NULL,Lee.12.JCP}
\begin{equation}
U=\mathrm{exp}[i\sigma_z\Omega/2],~~\Omega=2i\sum_{j}\frac{g_{j}}{\omega_{j}}(b^{\dagger}_{j}-b_{j})
\end{equation}
on the Hamiltonian Eq. (\ref{eq:h}) such that
\begin{eqnarray}
H_T &=& U^{\dagger}HU\nonumber\\
&=& H_0+H_I,
\end{eqnarray}
where the total free Hamiltonian is $H_0=H_s+H_B$ with the transformed system Hamiltonian reads
\begin{equation}
H_s=\frac{\varepsilon}{2}\sigma_z,
\end{equation}
and the bath Hamiltonian remains unaffected, $H_B=\sum_{j}\omega_{j}b^{\dagger}_{j}b_{j}$. The transformed interaction term takes the following form
\begin{equation}\label{eq:t_interaction}
H_I~=~\frac{\Delta}{2}(\sigma_x\cos\Omega+\sigma_y\sin\Omega).
\end{equation}
It's evident that $H_I$ is nonperturbative in the system-bath coupling strength.

Generally, the transformed interaction term is constructed such that $\langle H_I\rangle_{H_0}=0$ \cite{Lee.12.JCP}. By doing so, a perturbation theory in $H_I$ can be obtained. Notice the thermal average $\langle\sin\Omega\rangle_{H_0}=0$, Eq. (\ref{eq:t_interaction}) yields $\langle H_I\rangle_{H_0}=\frac{\Delta}{2}\xi\langle\sigma_x\rangle_{H_s}$ with \cite{Wang.15.SR}
\begin{equation}\label{eq:xi}
\xi~=~\exp\left[-\int_0^{\infty}d\omega\frac{J(\omega)}{\pi\omega^2}(n_B+\frac{1}{2})\right],
\end{equation}
where $n_B$ denotes the Bose-Einstein distribution. For the bath spectral function we choose [c.f., Eq. (\ref{eq:bath_spec})], it can be easily verified that for $s\leqslant 1$ the integral in Eq. (\ref{eq:xi}) becomes divergent, indicating the expectation of $H_I$ will always approach to zero regardless of the system-bath coupling strength. While for $s>1$, the integral is finite. Therefore, in the cases of Ohmic and sub-Ohmic dissipations, we can safely treat $H_I$ as a perturbation. For the super-Ohmic dissipation, we should further adopt a fluctuation-decoupling scheme \cite{Lee.12.JCP,Wang.15.SR} and choose $H_I-\langle H_I\rangle_{H_0}$ as the transformed interaction term. Thus after the PT, a strong system-bath interaction in $H$ is transformed into an effective weak system-bath interaction in $H_T$.

For convenience, we limit ourself to the Ohmic and sub-Ohmic dissipations in this study, the extension to the super-Ohmic dissipation is straightforward. Note that $[\sigma_z, U]=0$, thus the GFs of $\sigma_z$ is invariant with respect to the PT. We can evaluate them in the transformed Hamiltonian $H_T$.

\section{Evaluation of Green's Functions}\label{sec:3}
In this section, we will present details about the method we henceforth develop for GFs. Note that spin operators do not satisfy the Wick theorem. In order to overcome this difficulty, the so-called MFR \cite{Tsvelik.92.PRL} is utilized in the method such that standard Feynman diagram techniques as well as the Dyson's equation can be used.

\subsection{Majorana-fermion representation}
To make the present study self-sufficient, a brief note on the MFR is given
here. Technically, the methodology involves the introduction of a triplet of real fermions $\eta_{\alpha}$ (with $\alpha = x,y,z$) that satisfy \cite{Shnirman.03.PRL}
\begin{equation}
\eta_{\alpha}\eta_{\beta}~=~-\eta_{\beta}\eta_{\alpha}~~(\alpha\neq\beta),~~~\eta_{\alpha}^2=1,
\end{equation}
it leads to a representation of spin operators
\begin{equation}\label{eq:MF}
\sigma_{\alpha}=-i\sum_{\beta\gamma}\epsilon_{\alpha\beta\gamma}\eta_{\beta}\eta_{\gamma}.
\end{equation}

Noting the crucial property of the MFR
$\left\langle\sigma_{\alpha}(\tau)\sigma_{\beta}\right\rangle~=~\left\langle\eta_{\alpha}(\tau)\eta_{\beta}\right\rangle$,
if we introduce the GFs of Majorana-fermions
\begin{equation}\label{eq:mf_GF}
G_{\alpha\beta}(\tau)~\equiv~-i\left\langle T~\eta_{\alpha}(\tau)\eta_{\beta}\right\rangle,
\end{equation}
then GFs of spin operators can be rewritten in terms of the greater and lesser GF of Majorana-fermions, namely,
$\Pi_{\alpha\beta}^<(\tau) = -G_{\alpha\beta}^<(\tau),
\Pi_{\alpha\beta}^>(\tau) = G_{\alpha\beta}^>(\tau),
\Pi_{\alpha\beta}^r(\tau) = \Theta(\tau)[G_{\alpha\beta}^{>}(\tau)+G_{\alpha\beta}^{<}(\tau)],
\Pi_{\alpha\beta}^a(\tau) = -\Theta(-\tau)[G_{\alpha\beta}^{>}(\tau)+G_{\alpha\beta}^{<}(\tau)]$.
Thus the evaluation of GFs $\Pi_{\alpha\beta}$ turns into an evaluation of GFs $G_{\alpha\beta}$. The \SSCF{} now can be expressed as
\begin{equation}\label{eq:corr_w}
S_z(\omega) ~=~ \frac{i}{2}\left[G_{zz}^>(\omega)-G_{zz}^<(\omega)\right].
\end{equation}
For later convenience, we denote $G_{\eta}\equiv G_{zz}$, $G_{\eta_x} \equiv G_{xx}$, $G_{\eta_y} \equiv G_{yy}$.

\subsection{Free Green's functions}
In order to calculate Majorana-fermion GFs [Eq. (\ref{eq:mf_GF})] via the Dyson's equation, the first step is to obtain explicit forms for free GFs of fermions and bath operators.

\subsubsection{Majorana-fermion}
In the MFR, the transformed system Hamiltonian $H_s$ can be expressed as
\begin{equation}\label{eq:hs}
H_s~=~-i\frac{\varepsilon}{2}\eta_x\eta_y.
\end{equation}
For an isolated system described by Eq. (\ref{eq:hs}), equations of motion that $\eta_x$ and $\eta_y$ satisfy are given by
$\eta_y(t)=\eta_y\cos\varepsilon t+\eta_x\sin\varepsilon t$ and
$\eta_x(t)=\eta_x\cos\varepsilon t-\eta_y\sin\varepsilon t$, respectively. According to the definitions for GFs, we find greater and lesser components of the free GFs $G_{\eta_x,0}$ and $G_{\eta_y,0}$ in the Fourier space have the following forms
\begin{eqnarray}\label{eq:gxy}
G_{\eta_x,0}^>(\omega) &=& G_{\eta_y,0}^>(\omega) = -i2\pi\delta(\omega-\varepsilon),\nonumber\\
G_{\eta_x,0}^<(\omega) &=& G_{\eta_y,0}^<(\omega) = i2\pi\delta(\omega+\varepsilon).
\end{eqnarray}
In deriving the above equations, we choose a spin-down initial state such that $\left\langle\eta_x\eta_y\right\rangle=-i$, but we have checked that the final results is independent of initial conditions.

Note that $\eta_z$ commutes with $H_s$, so $\eta_z$ is time independent. We find
\begin{equation}
G_{\eta,0}^{r/a}(\tau)~=~\mp 2i \Theta(\pm \tau),
\end{equation}
in the frequency domain, we have
\begin{equation}\label{eq:geta0_ra}
G_{\eta,0}^{r/a}(\omega)~=~\frac{2}{\omega\pm i\xi},~~~~~\xi\rightarrow 0.
\end{equation}

\subsubsection{Bath}
For the bath operator, we introduce compact notations $B\equiv(\cos\Omega,\sin\Omega)^T$ and $B^{\dagger}\equiv(\cos\Omega,\sin\Omega)$ and define a matrix GF
\begin{equation}
G_{B,0}(\tau)\equiv-i\left\langle T B(\tau)B^{\dagger}\right\rangle.
\end{equation}
Noting relations between four components of the GF, in the following, we only evaluate the lesser and greater components of $G_{B,0}$, namely,
$G_{B,0}^>(\tau)= -i\left\langle B(\tau)B^{\dagger}\right\rangle$ and
$G_{B,0}^<(\tau)= -i\left\langle B^{\dagger}B(\tau)\right\rangle$.

In terms of correlation functions defined as $\Phi_{nm}(\tau)\equiv\frac{\Delta^2}{4}\left\langle e^{ni\Omega(\tau)}e^{mi\Omega}\right\rangle$ and $\tilde{\Phi}_{nm}(\tau)\equiv\frac{\Delta^2}{4}\left\langle e^{ni\Omega}e^{mi\Omega(\tau)}\right\rangle=\Phi_{nm}(-\tau)$ with $n,m=\pm$, we can rewrite matrix elements of $G_{B,0}^{>,<}$. For instance, we have
\begin{eqnarray}
\left\langle \cos\Omega(\tau)\cos\Omega\right\rangle &=& \frac{1}{\Delta^2}\left[\Phi_{++}(\tau)+\Phi_{+-}(\tau)\right.\nonumber\\
&&\left.+\Phi_{-+}(\tau)+\Phi_{--}(\tau)\right]
\end{eqnarray}
for an element of the greater GF, by replacing $\Phi_{nm}(\tau)$ with $\tilde{\Phi}_{nm}(\tau)$, we can obtain results for elements of the lesser GF.

Those correlation functions $\Phi_{nm}(\tau)$ and $\tilde{\Phi}_{nm}(\tau)$ can be evaluated by using the techniques of Feynman disentangling of operators \cite{Mahan.00.NULL}. Here, we write down the final results (details can be found in \ref{sec:cf_bath})
\begin{eqnarray}
\Phi_{+-}(\tau) &=& \Phi_{-+}(\tau)=\frac{\Delta^2}{4}\exp[-Q_2(\tau)-iQ_1(\tau)],\\\label{eq:phipm}
\tilde{\Phi}_{+-}(\tau) &=& \tilde{\Phi}_{-+}(\tau)=\frac{\Delta^2}{4}\exp[-Q_2(\tau)+iQ_1(\tau)]\label{eq:tphipm}
\end{eqnarray}
with
\begin{eqnarray}
Q_1(\tau) &=& \frac{2}{\pi}\int_0^{\infty}d\omega\frac{J(\omega)}{\omega^2}\sin\omega \tau,\label{eq:q1}\\
Q_2(\tau) &=& \frac{2}{\pi}\int_0^{\infty}d\omega\frac{J(\omega)}{\omega^2}\coth\left(\frac{\omega}{2T}\right)(1-\cos\omega \tau). \label{eq:q2}
\end{eqnarray}

With the spectral form Eq. (\ref{eq:bath_spec}), the integrals in the bath correlation functions can be evaluated exactly \cite{Gorlich.88.PRB}
\begin{eqnarray}
Q_2(\tau)+iQ_1(\tau) &=& 2\alpha\Gamma(s-1)\left[1-z^{1-s}\right]+2\alpha\Gamma(s-1)\kappa^{1-s}\left[2\zeta(s-1,1+1/\kappa)\right.\nonumber\\
&&\left.-\zeta(s-1,1+z/\kappa)-\zeta(s-1,1+z^{\ast}/\kappa)\right],\label{eq:phipm_exact}\\
Q_2(\tau)-iQ_1(\tau) &=& 2\alpha\Gamma(s-1)\left[1-z^{\ast 1-s}\right]+2\alpha\Gamma(s-1)\kappa^{1-s}\left[2\zeta(s-1,1+1/\kappa)\right.\nonumber\\
&&\left.-\zeta(s-1,1+z^{\ast}/\kappa)-\zeta(s-1,1+z/\kappa)\right],\label{eq:tphipm_exact}
\end{eqnarray}
where $\kappa=\omega_c/T$, $z=1+i\omega_c\tau$ and $\zeta(z,q)$ is Riemann's generalized zeta function.

Similarly, we find $\Phi_{++}(\tau)=\Phi_{--}(\tau)$ and $\tilde{\Phi}_{++}(\tau)=\tilde{\Phi}_{--}(\tau)$
(concrete forms are presented in \ref{sec:cf_bath}).
By noting, for instance, $\left\langle \cos\Omega(\tau)\sin\Omega\right\rangle = \frac{i}{\Delta^2}\left[\Phi_{+-}(\tau)-\Phi_{++}(\tau)+\Phi_{--}(\tau)-\Phi_{-+}(\tau)\right]$, we deduce that the nondiagonal elements of $G_{B,0}^{>,<}$, namely, $\left\langle \cos\Omega(\tau)\sin\Omega\right\rangle$ and $\left\langle \sin\Omega(\tau)\cos\Omega\right\rangle$ in $G_{B,0}^>$, $\left\langle \cos\Omega\sin\Omega(\tau)\right\rangle$ and $\left\langle \sin\Omega\cos\Omega(\tau)\right\rangle$ in $G_{B,0}^<$, are vanishing, so $G_{B,0}^{>,<}$ are actually diagonal matrices.

\subsection{Self-energies and Green's functions}
In the following, we first focus on the extraction of the fermionic self-energies, then utilize the Dyson's equation to obtain explicit forms for Majorana-fermion GFs. In the MFR, the interaction part can be rewritten as
\begin{equation}\label{eq:coupling}
H_I~\equiv~-i\frac{\Delta}{2}\left(\eta_y\eta_z\cos\Omega+\eta_z\eta_x\sin\Omega\right).
\end{equation}
As be demonstrated before, this interaction can safely be regarded as a perturbation. For such a weak interaction, the spin dynamics should, in principle, be well described by the lowest-order self-energies. The lowest nonvanishing correction to $G_{\eta,0}$ is
\begin{equation}
\frac{i}{2}\int dt_1\int dt_2\mathrm{Tr}\left\{\rho_0T\left[\eta_z(t)\eta_z(t^{\prime})H_I(t_1)H_I(t_2)\right]\right\}
\end{equation}
with $\rho_0$ as the initial total density matrix. In what follows, we choose a factorized initial condition, namely, $\rho_0=\rho_B\otimes\rho_s$. The reservoir is prepared in a canonical-equilibrium state of temperature $T\equiv\beta^{-1}$
\begin{equation}
\rho_B~=~\frac{e^{-\beta H_B}}{Z_B},~~~Z_B=\mathrm{Tr}e^{-\beta H_B}.
\end{equation}
The spin system $H_s$ is prepared in a spin-down state. From the above form, the leading order self-energy $\Sigma_{\eta}$ is obtained at a second order of tunneling $\Delta$ where both fermionic and bosonic
lines are given by the free propagators
\begin{eqnarray}\label{eq:se_t}
\Sigma_{\eta}(t_1,t_2) &=&  i\frac{\Delta^2}{4}\left[_{11}G_{B,0}(t_1,t_2)G_{\eta_y,0}(t_1,t_2)\right.\nonumber\\
&&\left.+_{22}G_{B,0}(t_1,t_2)G_{\eta_x,0}(t_1,t_2)\right]
\end{eqnarray}
with $_{11}G_{B,0}\equiv\left(\begin{array}{cc}
1 & 0
\end{array}
\right)G_{B,0}\left(\begin{array}{c}
1 \\
0
\end{array}
\right)$ and $_{22}G_{B,0}\equiv\left(\begin{array}{cc}
0 & 1
\end{array}
\right)G_{B,0}\left(\begin{array}{c}
0 \\
1
\end{array}
\right)$.
This level of approximation corresponds to the second order Born approximation.

Applying the Langreth theorem \cite{Haug.96.NULL} together with the above free Green functions to $\Sigma_{\eta}$, we easily find that
$\Sigma_{\eta}^>(\omega)=-i\Phi_{+-}(\omega-\varepsilon)$ and
$\Sigma_{\eta}^<(\omega)=i\tilde{\Phi}_{+-}(\omega+\varepsilon)$. Noting a relation between GFs $G^r-G^a=G^>-G^<$, we have
\begin{equation}
\Sigma_{\eta}^r(\omega)-\Sigma_{\eta}^a(\omega) = -i\left(\Phi_{+-}(\omega-\varepsilon)+\tilde{\Phi}_{+-}(\omega+\varepsilon)\right),
\end{equation}
from which we can directly obtain imaginary parts of $\Sigma_{\eta}^{r/a}$ ($\Sigma_{\eta}^{r\ast}=\Sigma_{\eta}^a$)
\begin{equation}\label{eq:eta_ra}
\mathrm{Im}\left[\Sigma_{\eta}^{r/a}(\omega)\right]~\equiv~\mp\frac{1}{2}\Gamma(\omega).
\end{equation}
with "$\mathrm{Im}$" the imaginary part and $\Gamma(\omega)\equiv\left[\Phi_{+-}(\omega-\varepsilon)+\tilde{\Phi}_{+-}(\omega+\varepsilon)\right]$.

As for the real part of the complex-conjugated retarded and advanced self-energies, it gives the Lamb shift, i.e., renormalize the level splitting. Usually, the real part of the self-energy is ignored in perturbation theories. However, in order to capture dissipative features of spin dynamics in the original strong-coupling regime where the renormalization of level splitting is obvious, we must take the real part into account in our theory. The real part $\Lambda$ of $\Sigma_{\eta}^{r/a}$ can be obtained as (see details in \ref{sec:re_se})
\begin{eqnarray}\label{eq:eta_ra_real}
\Lambda(\omega) &=& \frac{1}{2\pi}\mathcal{P}\int\frac{\left[\Phi_{+-}(\omega-\omega^{\prime})+\tilde{\Phi}_{+-}(\omega-\omega^{\prime})\right]\omega^{\prime}}{\omega^{\prime 2}-\varepsilon^2}d\omega^{\prime}\nonumber\\
&&-\frac{1}{2}\mathrm{Im}\left\{\left.\left[\Phi_{+-}(\lambda)-\tilde{\Phi}_{+-}(\lambda)\right]\right|_{\lambda=-i(\omega+\varepsilon)}\right\}\nonumber\\
&&+\frac{1}{2}\mathrm{Im}\left\{\left.\left[\Phi_{+-}(\lambda)-\tilde{\Phi}_{+-}(\lambda)\right]\right|_{\lambda=-i(\omega-\varepsilon)}\right\},
\end{eqnarray}
where $\mathcal{P}$ means the Cauchy principal value, $\Phi_{+-}(\lambda)$ and $\tilde{\Phi}_{+-}(\lambda)$ are bath correlations in the Laplace space. The sum of the retarded and advanced self-energies give the Keldysh component of self-energy
\begin{equation}
\Sigma_{\eta}^K(\omega) ~=~ i\left[\tilde{\Phi}_{+-}(\omega+\varepsilon)-\Phi_{+-}(\omega-\varepsilon)\right].
\end{equation}

Now we can obtain explicit forms for $G_{\eta}^{r/a/K}(\omega)$. According to the Dyson's equation $G^{-1}=G_0^{-1}-\Sigma$, we have
\begin{equation}\label{eq:geta_ra}
G_{\eta}^{r/a}(\omega)
~=~\frac{2}{\omega-2\Lambda\pm i\Gamma}.
\end{equation}
Inserting the above results into the Keldysh equation $G^K=G^r\Sigma^KG^a$, we find
\begin{equation}\label{eq:geta_k}
G_{\eta}^{K}(\omega)
~=~ \frac{4i \left[\tilde{\Phi}_{+-}(\omega+\varepsilon)-\Phi_{+-}(\omega-\varepsilon)\right]}{(\omega-2\Lambda)^2+ \Gamma^2}.
\end{equation}
Eqs. (\ref{eq:geta_ra}) and (\ref{eq:geta_k}) allow us, at least in the leading order, to use the relation $G_{\eta}^K(\omega)=h_{\eta}(\omega)[G_{\eta}^r(\omega)-G_{\eta}^a(\omega)]$, where the scalar
\begin{equation}
h_{\eta}(\omega)~=~-\frac{\tilde{\Phi}_{+-}(\omega+\varepsilon)-\Phi_{+-}(\omega-\varepsilon)}{\Gamma}.
\end{equation}

Using a relation $G^<=\frac{1}{2}(G^K-G^r+G^a)$, we obtain
\begin{equation}\label{eq:gzzl}
G_{\eta}^<(\omega)
~=~ \frac{4i \tilde{\Phi}_{+-}(\omega+\varepsilon)}{(\omega-2\Lambda)^2+ \Gamma^2}.
\end{equation}
Similarly, we have
\begin{equation}\label{eq:gzzg}
G_{\eta}^>(\omega)
~=~ -\frac{4i \Phi_{+-}(\omega-\varepsilon)}{(\omega-2\Lambda)^2+ \Gamma^2}.
\end{equation}
Noting the mapping between components of spin GFs $\Pi$ and Majorana-fermion GFs $G$, various GFs of the spin operator $\sigma_z$ can be determined.

\subsection{Observables}
Now, with the acquired knowledge of GFs, we can evaluate quantities of interest. Inserting Eqs. (\ref{eq:gzzl}) and (\ref{eq:gzzg}) into Eq. (\ref{eq:corr_w}), we arrive at a formal form for the \SSCF{} in the Fourier space
\begin{eqnarray}\label{eq:sz_fourier}
S_z(\omega)
&=& \frac{2\Gamma}{(\omega-2\Lambda)^2+ \Gamma^2}\nonumber\\
&=& 2\mathrm{Re}\frac{1}{-i\omega+2i\Lambda+\Gamma},
\end{eqnarray}
where "$\mathrm{Re}$" denotes the real part. We can rewrite $S_z(\omega)=2\mathrm{Re}\tilde{S}_z(\lambda=-i\omega)$ by introducing the \SSCF{} in the Laplace space
\begin{equation}\label{eq:sz_b}
\tilde{S}_z(\lambda)
~=~\frac{1}{\lambda+g_z(\lambda)}
\end{equation}
with the kernel satisfies (details can be found in \ref{sec:kernel})
\begin{equation}\label{eq:sz_kernel}
g_z(\lambda)~=~2\tilde{\Phi}_{+-}(\lambda-i\varepsilon)+2\Phi_{+-}(\lambda+i\varepsilon).
\end{equation}
Both forms of the \SSCF{} [c.f., Eqs. (\ref{eq:sz_fourier}) and (\ref{eq:sz_b})] will be frequently used in the following investigations. Specifically, Eq. (\ref{eq:sz_fourier}) is appropriate for illustrating the dynamical features of the dissipative system, while Eq. (\ref{eq:sz_b}) is suitable for analytical treatments. According to Eqs. (\ref{eq:d_sus}) and (\ref{eq:s_sus}), we can further study susceptibilities through the \SSCF{}.

\section{Ohmic dissipation}\label{sec:4}
We first consider the Ohmic dissipation. In the scaling limit, we can take $\kappa\rightarrow\infty$, the bath correlation functions [c.f., Eqs. (\ref{eq:phipm}) and (\ref{eq:tphipm})] reduce to \cite{Dattagupta.89.JPCM,Weiss.12.NULL,Gorlich.88.PRB}
\begin{eqnarray}
\Phi_{+-}(\tau) &=& \frac{\Delta^2}{4}\mathrm{exp}\left[-i\pi\alpha\, \mathrm{sgn}(\tau)\right]\left(\frac{\pi T}{\omega_c\left[\sinh(\pi T|\tau|)\right]}\right)^{2\alpha},\\
\tilde{\Phi}_{+-}(\tau) &=& \frac{\Delta^2}{4}\mathrm{exp}\left[i\pi\alpha \,\mathrm{sgn}(\tau)\right]\left(\frac{\pi T}{\omega_c\left[\sinh(\pi T|\tau|)\right]}\right)^{2\alpha},
\end{eqnarray}
where $\mathrm{sgn}(x)$ denotes the sign function of $x$. By performing the Laplace transform, we get
\begin{eqnarray}\label{eq:bc_s}
\Phi_{+-}(\lambda) &=& \frac{\Delta^2}{4\omega_c}\mathrm{exp}\left(-i\pi\alpha\right)\left(\frac{2\pi T}{\omega_c}\right)^{2\alpha-1}\frac{\Gamma(1-2\alpha)\Gamma(\alpha+\lambda/2\pi T)}{\Gamma(1-\alpha+\lambda/2\pi T)},\nonumber\\
\tilde{\Phi}_{+-}(\lambda)&=& \frac{\Delta^2}{4\omega_c}\mathrm{exp}\left(i\pi\alpha\right)\left(\frac{2\pi T}{\omega_c}\right)^{2\alpha-1}\frac{\Gamma(1-2\alpha)\Gamma(\alpha+\lambda/2\pi T)}{\Gamma(1-\alpha+\lambda/2\pi T)}.
\end{eqnarray}

\subsection{Unbiased system}
For $\varepsilon=0$, the kernel [c.f. Eq. (\ref{eq:sz_kernel})] reduces to
\begin{equation}\label{eq:kernel_unbias}
g_z(\lambda)
~=~ \Delta_{e}\left(\frac{\Delta_{e}}{2\pi T}\right)^{1-2\alpha}\frac{\Gamma(\alpha+\lambda/2\pi T)}{\Gamma(1-\alpha+\lambda/2\pi T)},
\end{equation}
where we have utilized Eq. (\ref{eq:bc_s}) and the effective tunneling defined by \cite{Weiss.12.NULL}
\begin{equation}
\Delta_{e}=[\Gamma(1-2\alpha)\cos(\pi\alpha)]^{1/2(1-\alpha)}(\Delta/\omega_c)^{\alpha/(1-\alpha)}\Delta.
\end{equation}
We thus found the \SSCF{} [Eq. (\ref{eq:sz_b})] with the kernel [Eq. (\ref{eq:kernel_unbias})] coincides with the result obtained by the NIBA \cite{Weiss.12.NULL}. Although being so distinct in methodologies of these two approach, such an agreement can be understood by noting an equivalence between the NIBA and the general relaxation theory with the second-order Born approximation in unbiased systems \cite{Aslangul.86.JP,Dekker.87.PRA}. Compared our theory with the latter theoretical scheme, we found that the level of approximation is the same, namely, both theories have utilized the PT and invoked the second order Born approximation in perturbation expansions.

In the weak coupling regime of $\alpha\ll 1$, analytically, one only needs to keep the leading order of the kernel [Eq. (\ref{eq:kernel_unbias})] as $O(\alpha)$ such that
\begin{equation}
g_z(\lambda) ~\simeq~ \frac{\Delta^2}{\lambda}\left\{1+2\alpha\left[\ln\frac{2\pi T}{\Delta}+\frac{\pi T}{\lambda}+\psi\left(\frac{\lambda}{2\pi T}\right)\right]\right\}
\end{equation}
with $\psi(z)$ the digamma function. In deriving the above result, we have used a fact that $\Delta_e\approx\Delta$ in the limit of $\alpha\ll1$. Since $g_z(\lambda=-i\omega)=2i\Lambda(\omega)+\Gamma(\omega)$, we find $\Gamma(\omega)=\frac{\Delta^2}{\omega}\pi\alpha\coth\left(\frac{\omega}{2T}\right)$ and $2\Lambda(\omega)=\frac{\Delta^2}{\omega}\left\{1+2\alpha\left[\mathrm{Re}\psi\left(\frac{i\omega}{2\pi T}\right)+\ln\frac{2\pi T}{\Delta}\right]\right\}$. For weak noise spectra, the Lamb shift is weak, so we can ignore terms of order $\alpha$ in $\Lambda$. According to Eq. (\ref{eq:sz_fourier}), we have
\begin{equation}
S_z(\omega)~=~\frac{2\Delta^2\pi\alpha\coth\left(\frac{\omega}{2T}\right)}{(\omega^2-\Delta^2)^2+\Delta^2\left(\pi\alpha\coth\left(\frac{\omega}{2T}\right)\right)^2}.
\end{equation}
A previous result obtained by a perturbation theory utilizing the MFR \cite{Shnirman.03.PRL,Yang.14.EL} is thus recovered as a weak coupling limit of the \SSCF{} obtained here.

\begin{figure}[tbh]
  \centering
  \includegraphics[width=0.5\columnwidth]{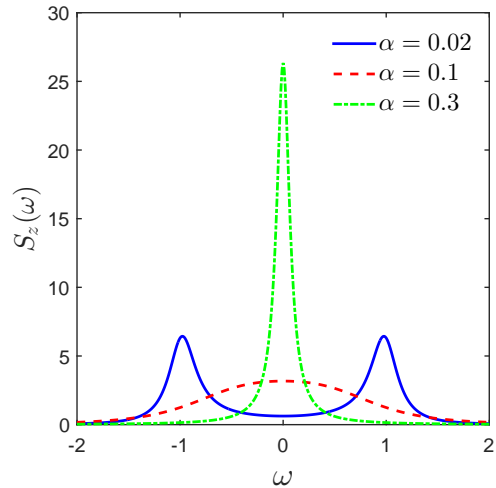}
\caption{(Color Online)Behaviors of \SSCF{} at $T=2.5$ with varying $\alpha$. The solid blue line denotes $\alpha=0.02$, the dashed red line denotes $\alpha=0.1$, the dashed-dotted green line denotes $\alpha=0.3$. Other parameters are $\Delta=1$, $\omega_c=30$.}
\label{fig:uo}
\end{figure}

General properties of the \SSCF{} can be found in \cite{Weiss.12.NULL}. But for the sake of completeness and clarity, we plot typical results for the \SSCF{} in the Fig. \ref{fig:uo}.
For finite temperatures, a dynamical crossover occurs in the range $0<\alpha<1/2$ with a transition temperature $T^{\ast}$ reads \cite{Weiss.12.NULL,Weiss.86.EL}
\begin{equation}\label{eq:transition_T} T^{\ast}=\frac{\Delta_e}{2\pi}\left\{\frac{h(0)}{\alpha^2}\left[1+\pi\alpha\cot(\pi\alpha)+2\sqrt{W(\alpha)}\right]\right\}^{\frac{1}{2(1-\alpha)}},
\end{equation}
where  $h(0)=\frac{\Gamma(1+\alpha)}{\Gamma(1-\alpha)}$, $W(\alpha)=\pi\alpha\cot(\pi\alpha)-\alpha^2g_1$, $g_1=\frac{1}{2}\left[\psi^{\prime}(1-\alpha)-\psi^{\prime}(1+\alpha)-g_2^2\right]$ with $\psi^{\prime}(z)$ the polygamma function of order 1 and $g_2=\frac{1}{\alpha}-\pi\cot(\pi\alpha)$. According to Eq. (\ref{eq:transition_T}), we know that for $\alpha=0.02, 0.1$ and $0.3$ with $\omega_c=30$ and $\Delta=1$, the transition temperature $T^{\ast}=16.3, 2.99$ and $0.45$, respectively.  Thus increasing the coupling strength in the Fig. \ref{fig:uo}, we see the \SSCF{} displays a transition behavior, namely, two peaks move towards the origin and finally turn into a narrow single peak.

\begin{figure}[tbh]
  \centering
  \includegraphics[width=0.5\columnwidth]{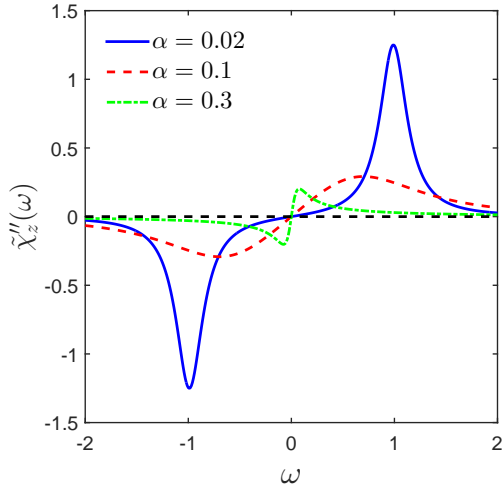}
\caption{(Color Online)Imaginary part of the dynamical susceptibility $\tilde{\chi}_z^{\prime\prime}(\omega)$ at $T=2.5$ with varying $\alpha$. The solid blue line denotes $\alpha=0.02$, the dashed red line denotes $\alpha=0.1$, the dashed-dotted green line denotes $\alpha=0.3$. Other parameters are $\Delta=1$, $\omega_c=30$.}
\label{fig:sus_uo}
\end{figure}

Effect of temperature on the excitation spectrum of the spin can be extracted from the imaginary part of the dynamical susceptibility, namely, $\tilde{\chi}_z^{\prime\prime}(\omega)$. According to the Fluctuation-dissipation theorem [c.f. Eq. (\ref{eq:d_sus})], $\tilde{\chi}_z^{\prime\prime}(\omega)$ is totally determined by the \SSCF{}, results are shown in the Fig. \ref{fig:sus_uo}.
As can be seen from the figure, for weak dissipations, the spectrum has two distinct peaks, corresponding to tunneling processes between two energy levels. For intermediate dissipations, the peaks broaden and move towards the origin, implying that the effective tunneling decreases as dissipation increases. When the system enters into the incoherent regime, the spectrum is confined to a region around $\omega=0$, the spin hardly flips.

\subsection{Biased system}
Theoretically, this case has been mostly described within the NIBA \cite{Leggett.87.RMP,Weiss.12.NULL,Weiss.89.PRL,Goerlich.89.EL,Weiss.86.EL,Grabert.85.PRL,Fisher.85.PRL}. It has been shown that the quantum tunneling rate has an extra power-law dependence on the bias \cite{Grabert.85.PRL,Fisher.85.PRL}. A quasi-elastic peak located at the origin induced by the bias in the spin correlation functions was also found \cite{Weiss.12.NULL}. However, the NIBA breaks down in the parameter region in which the system exhibits underdamped oscillations \cite{Leggett.87.RMP,Weiss.89.PRL}. It is then of necessity to verify the validity of our theory in this case.

\begin{figure}[tbh]
  \centering
  \includegraphics[width=0.75\columnwidth]{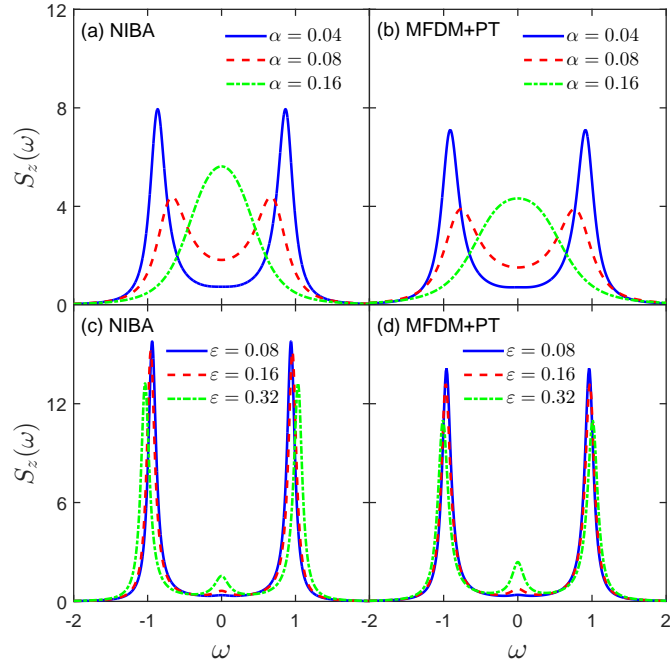}
\caption{(Color Online)Theoretical predictions on the \SSCF{}: a comparison between the NIBA [Refs. \cite{Dattagupta.89.JPCM,Weiss.12.NULL}] (left panel) and the present "MFDM+PT" [Eq. (\ref{eq:sz_b})] (right panel) at $T=1$ for (a)(b)$\varepsilon=0.1$ with varying coupling strength, (c)(d)$\alpha=0.02$ with varying bias. Other parameters are $\Delta=1$, $\omega_c=30$.}
\label{fig:biased_hT}
\end{figure}

We should mention that the \SSCF{} [c.f. Eq. (\ref{eq:sz_b})] is now different from the NIBA's prediction on Eq. (\ref{eq:corr_t}) \cite{Dattagupta.89.JPCM,Weiss.12.NULL}. We first focus on the high temperature regime $T\gtrsim\sqrt{\Delta_e^2+\varepsilon^2}$ in which the NIBA gives consistent results for spin correlation functions. The comparison at $T=1$ is shown in Fig. \ref{fig:biased_hT}. We choose parameters such that the high temperature requirement is always fulfilled. As can be seen, the line shapes obtained via two methods are in substantial agreement, either for fixed $\varepsilon$ with varying $\alpha$ or for fixed $\alpha$ with varying $\varepsilon$. For the former case [Fig. \ref{fig:biased_hT} (a)(b)], we find two inelastic peaks broaden and move towards the origin as the coupling strength increases, similar to the unbiased systems. The weight of the quasi-elastic peak induced by the bias is negligible compared with inelastic peaks' in the parameter regime we choose, so we hardly see it. For the latter case as shown in Fig. \ref{fig:biased_hT} (c)(d), we choose a weak coupling strength such that the Lamb shift of inelastic peaks is small. We can clearly observe a quasi-elastic peak located at the origin, its weight becomes larger as the bias increases. We also note that the quasi-elastic peak is more obvious in our result than in the NIBA's prediction. Since the NIBA omits initial system-bath correlations in obtaining the spin correlation functions, leading to an incorrect long-time behavior \cite{Sassetti.90.PRA,Sassetti.90.PRL}, consequently, the signal at $\omega=0$ will be affected.

\begin{figure}[tbh]
  \centering
  \includegraphics[width=0.75\columnwidth]{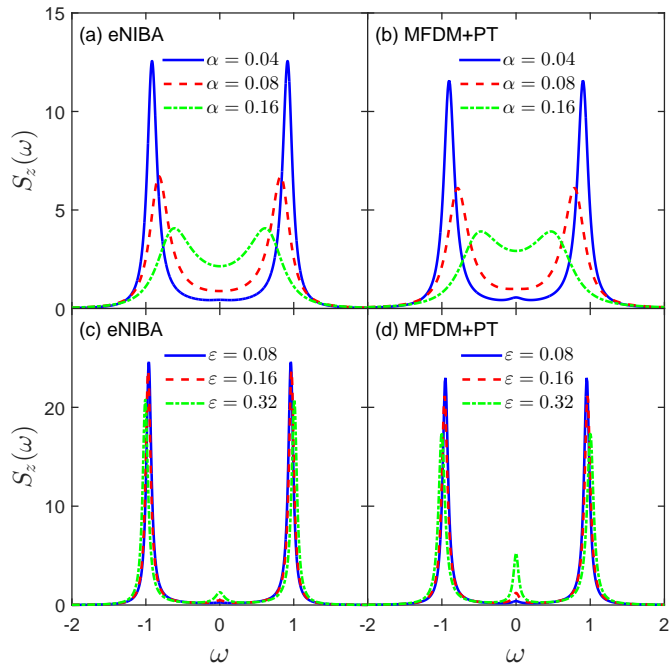}
\caption{(Color Online)Theoretical predictions on the \SSCF{}: a comparison between the eNIBA [Ref. \cite{Weiss.89.PRL}] (left panel) and the present "MFDM+PT" [Eq. (\ref{eq:sz_b})] (right panel) at $T=0.5$ for (a)(b)$\varepsilon=0.1$ with varying coupling strength, (c)(d)$\alpha=0.02$ with varying bias. Other parameters are $\Delta=1$, $\omega_c=30$.}
\label{fig:biased_lT}
\end{figure}

We then turn to the low temperature regime with $T<\sqrt{\Delta_e^2+\varepsilon^2}$. For comparison, we consider an extended-NIBA (eNIBA) which had taken the interblip interactions ignored in the original NIBA into account \cite{Weiss.89.PRL}. Results for the spin correlation function [Eq. (\ref{eq:corr_t})] at $T=0.5$ are shown in Fig. \ref{fig:biased_lT}. This value of temperature is chosen such that $\sqrt{\Delta_e^2+\varepsilon^2}$ is always bigger than it. The line shapes are again in substantial agreement, which means that our theory can be directly applied to low temperature regime of biased systems without any further modification, in contrast to the NIBA. The difference appears in the quasi-elastic peaks [Fig. \ref{fig:biased_lT} (c)(d)] is still related to the omission of initial system-bath correlations in the eNIBA.

\begin{figure}[tbh]
  \centering
  \includegraphics[width=0.75\columnwidth]{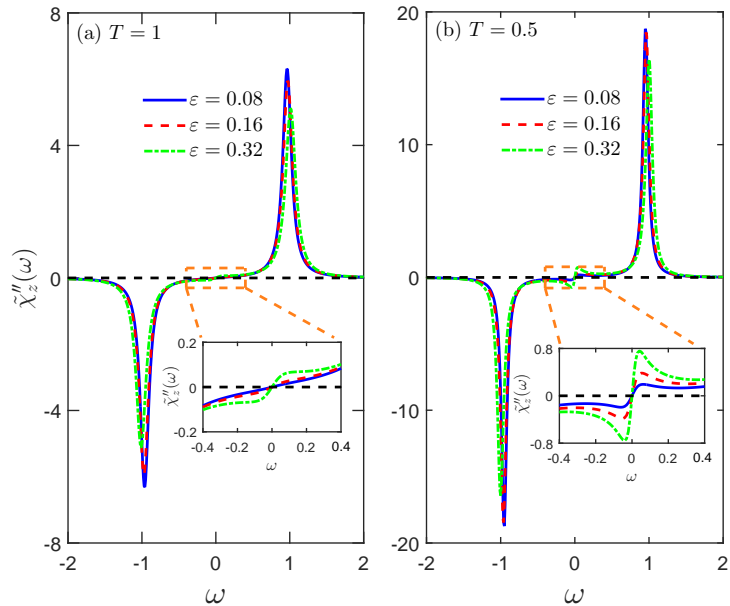}
\caption{(Color Online)Imaginary part of the dynamical susceptibility $\tilde{\chi}_z^{\prime\prime}(\omega)$ for (a)$T=1$, (b)$T=0.5$, with varying bias. The insets show the behaviors of $\tilde{\chi}_z^{\prime\prime}(\omega)$ near $\omega=0$. Other parameters are $\alpha=0.02$, $\Delta=1$, $\omega_c=30$.}
\label{fig:biased_sus}
\end{figure}

Finally we use the definition of $\tilde{\chi}_z^{\prime\prime}(\omega)$, namely, $-\mathrm{Im}\Pi_{zz}^r(\omega)$ or equivalently $\left[\Pi_{zz}^<(\omega)-\Pi_{zz}^>(\omega)\right]/2i$, to investigate its behaviors. Here, we focus on the effect of bias on the excitation spectrum of spin, thus we fix the coupling strength and vary the bias. In order to highlight the role of the bias, we choose a weak coupling strength, namely, $\alpha=0.02$. Results for both high and low temperature regime ($T=1$ and $T=0.5$, respectively) are summarized in Fig. \ref{fig:biased_sus}. We can see from the insets that step structures emerge at the origin, in contrast to unbiased systems in the weak coupling regime. Those structures are apparently induced by the bias, since the signal is enhanced as the bias increases. Besides the region around $\omega=0$, the spectrums are similar to those of the unbiased systems.

\section{Sub-Ohmic dissipation}\label{sec:5}
Compared with the Ohmic bath, low-frequency bath modes are more strongly pronounced in the sub-Ohmic case. Those bath modes lead to some peculiar and unexpected phenomena in the SBM, including a ultraslow quantum dynamics  \cite{Nalbach.10.PRB} and the persistence of coherent dynamics at strong dissipations \cite{Kast.13.PRL}. Besides the dynamics, its phase transition has also been investigated in detail recently. A continuous quantum phase transition (QPT) is found in the entire range $0<s<1$ by using the numerical renormalization-group (NRG) method \cite{Bulla.03.PRL}. On general grounds, one expect the QPT in the sub-Ohmic SBM to fall into the same universality class as that of the one-dimensional Ising chain with long-range interactions falling off as $r^{-s-1}$ \cite{Suzuki.76.PTP}. Noting the upper critical dimension of such an Ising chain is $2s$ \cite{Fisher.72.PRL,Aizenman.88.LMP}, thus the SBM with $0<s<1/2$ and $1/2<s<1$ correspond to the classical system falling above and below its upper critical dimension, respectively. As we will see, these two regimes depict distinct behaviors. So far, the quantum-to-classical mapping has been confirmed in the range $0<s<1/2$ by studying behaviors of the critical exponent $\nu$ and the dynamical exponent $z$ \cite{Bulla.03.PRL}. However, in the sub-Ohmic regime with $1/2<s<1$, no consensus has reached  \cite{Florens.11.PRB,Winter.09.PRL,Alvermann.09.PRL,Zhang.10.PRB,Chin.11.PRL,Guo.12.PRL,Kirchner.09.PRL,Kirchner.12.PRB}.

In the following, we limit ourself to unbiased systems, since the low-frequency modes generate an effective bias \cite{Nalbach.10.PRB}. We will see that the effective bias yields a quasi-elastic peak in the \SSCF{}, similar to the situation in the biased Ohmic SBM.

We first focus on the sub-Ohmic SBM at zero temperature. In this situation, Eq. (\ref{eq:phipm_exact}) reduces to $2\alpha\Gamma(s-1)\left[1-z^{1-s}\right]$, from which we obtain explicit forms for bath correlations $Q_1$ and $Q_2$ [Eqs. (\ref{eq:q1}) and (\ref{eq:q2})], namely, $Q_1(\tau) = -2\alpha\Gamma(s-1)\mathrm{Im}z^{1-s}$ and $Q_2(\tau) = 2\alpha\Gamma(s-1)\left[1-\mathrm{Re}z^{1-s}\right]$. Both functions are positive and monotonically increase in time, this leads to a damped oscillation in the kernel $g_z(\tau)$ by noting
$g_z(\tau)=\Delta^2e^{-Q_2(\tau)}\cos[Q_1(\tau)]$.

\begin{figure}[tbh]
  \centering
  \includegraphics[width=0.75\columnwidth]{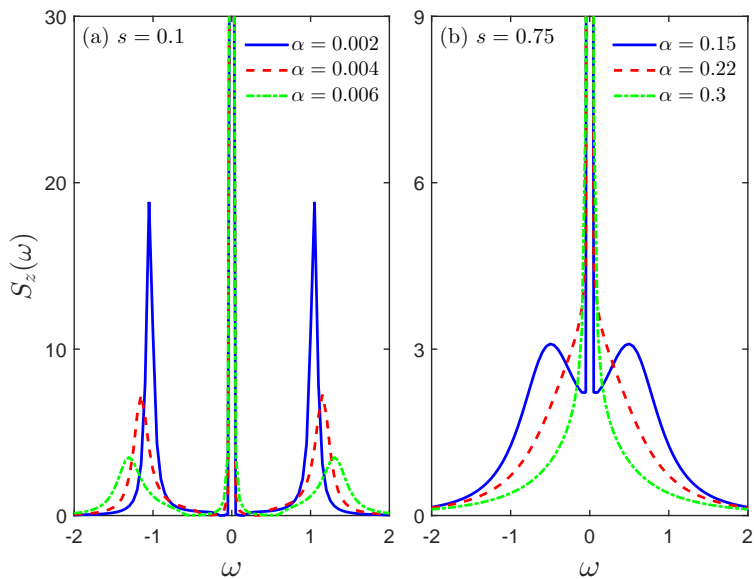}
\caption{(Color Online)The \SSCF{} of the unbiased sub-Ohmic SBM at $T=0$ for (a)$s=0.1$, (b)$s=0.75$, with varying coupling strength. Other parameters are $\Delta=1$, $\omega_c=10$.}
\label{fig:sub_zeroT}
\end{figure}

Using the expression for $g_z$, we can investigate the \SSCF{} at zero temperature. As we mentioned before, SBM with $0<s<1/2$ and $1/2<s<1$ depict different behaviors, so we fix $s$ and focus on the dependence of the \SSCF{} on the coupling strength. Typical results are shown in Fig. \ref{fig:sub_zeroT}.
Due to a bath-induced effective bias in the system, a quasi-elastic peak emerges in the \SSCF{} for unbiased systems. The weight of the peak is almost independent of the coupling strength since the effective bias is mainly affected by density of states of the low frequency bath modes. As for inelastic peaks, we observe totally different behaviors for small and large $s$ with varying the coupling strength. For $s=0.1$, two peaks move away from the origin with increasing $\alpha$, implying an absence of the dynamical crossover. This phenomenon is in accordance with the finding in Ref. \cite{Kast.13.PRL}. For $s=0.75$, two inelastic peaks move towards the origin with increasing $\alpha$ and finally turn into a single peak. This is the signature of the dynamical crossover. Actually, for $1/2<s<1$, the SBM exhibits a dynamical crossover at zero temperature \cite{Kast.13.PRL}.

We then turn to the quantum criticality of the static susceptibility, with an emphasis on the validity of the quantum-to-classical mapping. Recalled that for the associated long-range one-dimensional Ising chain falls above or below its upper critical dimension, the critical exponent $\gamma$ assumes a mean-field value $1$ or satisfies \cite{Fisher.72.PRL}
\begin{equation}\label{eq:gamma_critical}
\gamma^{-1}=1-\frac{1}{3}\frac{2s-1}{s}-\frac{1}{9}G(s)\left(\frac{2s-1}{s}\right)^2
\end{equation}
with $G(s)=s\left[\psi(1)-2\psi(s/2)+\psi(s)\right]$ and $\psi(z)$ the digamma function for $0<s<1/2$ or $1/2<s<1$, respectively. Recent numerical studies \cite{Winter.09.PRL,Alvermann.09.PRL,Zhang.10.PRB,Chin.11.PRL,Guo.12.PRL} have confirmed that $\gamma$ indeed follows the predicted mean-field value in the regime $0<s<1/2$ of the SBM. However, the validity of Eq. (\ref{eq:gamma_critical}) in the regime $1/2<s<1$ remains untouch so far.

\begin{figure}[tbh]
  \centering
  \includegraphics[width=0.7\columnwidth]{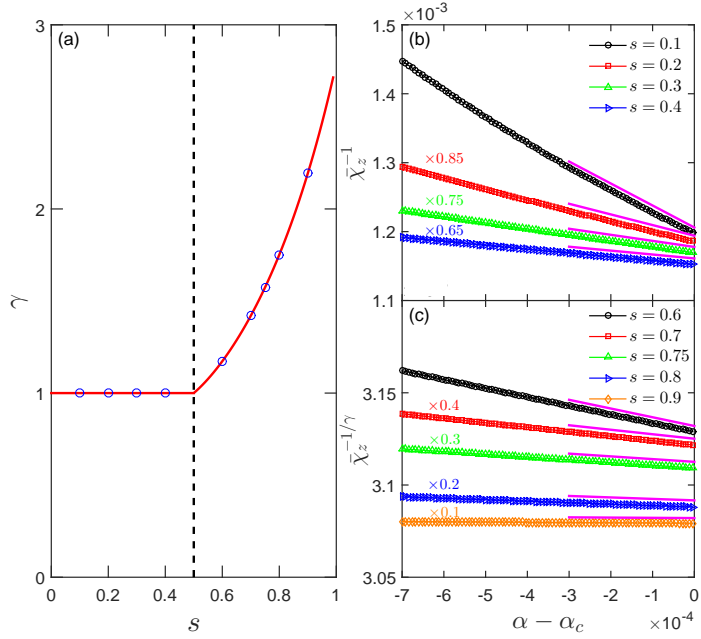}
\caption{(Color Online)Criticality of the static susceptibility of the unbiased sub-Ohmic SBM at $T=0$. (a)The critical exponent $\gamma$ obtained for the Ising chain with long-range interactions $r^{-1-s}$ [Eq. (\ref{eq:gamma_critical}) for $1/2<s<1$]. The circles indicate the parameter values which are used in the right panel. (b)Dependence of $\bar{\chi}_z^{-1}$ on $\alpha-\alpha_c$ in the regime $0<s<1/2$, the mean-field value $\gamma=1$ has been adopted. (c)Dependence of $\bar{\chi}_z^{-1/\gamma}$ on $\alpha-\alpha_c$ in the regime $1/2<s<1$. In (b) and (c), the curves for $\bar{\chi}_z$ are multiplied with the indicated factors for better visibility. The straight lines indicate the linear dependence in the vicinity of the quantum critical point $\alpha_c$ whose values are taken from Ref. \cite{Alvermann.09.PRL}.  Other parameters are $\Delta=1$, $\omega_c=10$.}
\label{fig:static_sus}
\end{figure}

We note that with the information of the \SSCF{}, Eq. (\ref{eq:s_sus}) enables us to investigate properties of the static susceptibility. Instead of directly determining the critical exponent from the data of the static susceptibility, we verify the linear dependence of $\bar{\chi}_z^{-1/\gamma}$ on $\alpha-\alpha_c$ by using the predictions of $\gamma$ for the Ising chain [Eq. (\ref{eq:gamma_critical})].  The result is summarized in the Fig. \ref{fig:static_sus}.
We calculate $\bar{\chi}_z$ by using Eq. (\ref{eq:s_sus}) for discrete values of $\alpha$. As can be seen from Figs. \ref{fig:static_sus}(b) and (c), the linear dependence of $\bar{\chi}_z^{-1/\gamma}$ on $\alpha-\alpha_c$ is apparent, implying the quantum-to-classical mapping holds in the entire range of $0<s<1$.

\begin{figure}[tbh]
  \centering
  \includegraphics[width=0.7\columnwidth]{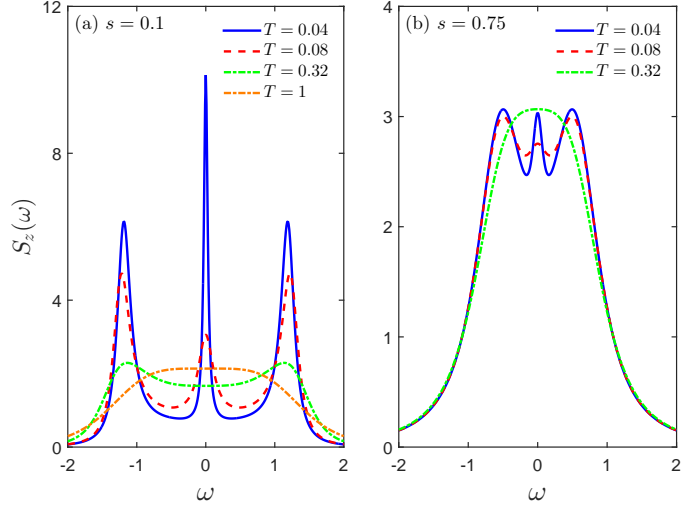}
\caption{(Color Online)The \SSCF{} of the unbiased sub-Ohmic SBM at finite temperatures for (a)$s=0.1$ with $\alpha=0.004$ and (b)$s=0.75$ with $\alpha=0.15$. Other parameters are $\Delta=1$, $\omega_c=10$.}
\label{fig:sub_T}
\end{figure}

We now move to the effect of finite temperatures on the dynamical crossover. According to Eq. (\ref{eq:phipm_exact}), a real part account for finite temperatures is added to $Q_2$, which leads to an exponential suppression of the kernel $g_z(\tau)$ and has thus the tendency to destroy coherence. Therefore, we expect in the regime $0<s<1/2$ that a dynamical crossover should occur at finite temperatures. Typical results for the \SSCF{} are presented in Fig. \ref{fig:sub_T}. We see from the Fig. \ref{fig:sub_T} (a) for $s=0.1$ that a dynamical crossover really happens. More interestingly, we found for both small and large $s$ that the quasi-elastic peak is more sensitive to the temperature change than inelastic peaks in the low temperature regime. Because in that regime, thermal fluctuations suppress the influence of the effective bias but are not enough to affect the spin tunneling process.

\section{Summary}\label{sec:6}
Methodologically, we established a theoretical approach for the calculation of Green¡¯s functions for spin-boson models in its complete generality. The theory involves two strategies, one is the Majorana-fermion representation such that standard Feynman diagram techniques as well as the Dyson's equation can be applied to spin systems, the other is the polaron transformation which enables us to investigate the strong system-bath coupling regime. Within this theoretical scheme, we can derive expressions for Green's functions that are valid over wide regimes of the coupling strength, bias as well as temperature, in contrast to conventional perturbation theories.

To demonstrate the utility of the approach, we consider the symmetrized spin correlation function (\SSCF{}) and the susceptibility in the spin-boson models. The \SSCF{} is related to the sum of the greater and lesser Green's function, thus an explicit expression can be obtained from Green's functions. Surprisingly, we found for the unbiased spin-boson models that the so-obtained expression for the \SSCF{} is identical to the result obtained via the noninteracting-blip approximation (NIBA). We attribute this coincidence to the fact that the NIBA actually bears resemblance to our theory by noting Refs. \cite{Aslangul.86.JP,Dekker.87.PRA}. In the weak coupling limit, we found the \SSCF{} reduces to a previous result obtained from a Majorana-fermion diagrammatic method \cite{Shnirman.03.PRL}. Furthermore, for biased spin-boson models, we confirm the validity of the \SSCF{} over wide regimes of temperature , it is highly in contrast to the NIBA which is not reliable in biased spin-boson systems at low temperatures.

As for the susceptibility, we utilize its relations with the \SSCF{} to carry on the investigations. For systems embedded in an Ohmic bath, we focus on the imaginary part of the dynamical susceptibility which gives the excitation spectrum of the spin. Effects of temperature, coupling strength as well as bias on the excitation spectrum are discussed. For systems embedded in a sub-Ohmic bath, we pay attention to the static susceptibility since its quantum critical behavior is of primary interest in recent years. We confirm the validity of the critical exponent $\gamma$ obtained for the one-dimensional Ising chain with long-range interactions in the sub-Ohmic spin-boson models, indicating the quantum-to-classical mapping holds in the whole range $0<s<1$.

Furthermore, this framework can be easily extended to a nonequilibrium case and thus enables us to study the heat current in spin-boson models from a unified perspective \cite{Liu.15.NULL}.

\section*{Acknowledgments}
The authors thank J. Cao, H. Zhou and J. Ren for highly useful discussions. Support from the National Basic Research Program of China with Grant No. 2012CB921401 is gratefully acknowledged. The work is also supported by the National Nature Science Foundation of China.

\appendix
\section{Bath correlation functions}\label{sec:cf_bath}
In this part, we use the techniques of Feynman disentangling of operators \cite{Mahan.00.NULL} to evaluate bath correlation functions. As an illustration, we calculate $\Phi_{-+}(\tau)$. According to the definition, we have
\begin{eqnarray}\label{eq:f}
\langle e^{-i\Omega (\tau)}e^{i\Omega}\rangle &=& \prod_{j}(1-e^{-\beta\omega_{j}})\sum_{n_{j}=0}^{\infty}e^{-\beta n_{j}\omega_{j}}\nonumber\\
&&\langle n_{j}|e^{\mu_{j}P_{j}(\tau)}e^{-\mu_{j}P_{j}}|n_{j}\rangle\nonumber\\
&\equiv& \prod_{j} \mathcal{F}_{j}(\tau)
\end{eqnarray}
with $\mu_{j}=2g_{j}/\omega_{j}$ and $P_{j}(\tau)=b_{j}^{\dagger}e^{i\omega_{j}\tau}-b_{j}e^{-i\omega_{j}\tau}$. Using the theorem $e^{A+B}=e^Ae^Be^{-1/2[A,B]}$, we find
\begin{equation}
e^{\mu_{j}P_{j}(\tau)}e^{-\mu_{j}P_{j}}~=~e^{-\mu_{j}^2}e^{\mu_{j}b_{j}^{\dagger}e^{i\omega_{j}\tau}}e^{-\mu_{j}b_{j}e^{-i\omega_{j}\tau}}e^{-\mu_{j}b_{j}^{\dagger}}e^{\mu_{j}b_{j}}.
\end{equation}
It's convenient to consider the normal ordering of operators, namely, all creation operators are to the left of all annihilation operators in the product. Then the center two operators need to be exchanged (for simplicity, we drop all subscripts)
\begin{eqnarray}
e^{-\mu b(\tau)}e^{-\mu b^{\dagger}} &=& e^{-\mu b^{\dagger}}e^{\mu^2 e^{-i\omega \tau}}e^{-\mu be^{-i\omega \tau}},
\end{eqnarray}
we have use the results $e^{\mu b^{\dagger}}be^{-\mu b^{\dagger}}=b-\mu$ and $e^{\mu b^{\dagger}}e^{-\mu b(\tau)}e^{-\mu b^{\dagger}}=\mathrm{exp}[-\mu e^{-i\omega \tau}(b-\eta)]$.

So $\mathcal{F}_{j}$ [Eq. (\ref{eq:f})] is finally arranged in to the desired form
\begin{eqnarray}
\mathcal{F}_{j}(\tau)
&=& e^{-\mu_{j}^2(1-e^{-i
\omega_{j}\tau})}(1-e^{-\beta\omega_{j}})\nonumber\\
&&\sum_{n_{j}}e^{-\beta\omega_{j}n_{j}}\sum_{n_{j}}\sum_{l=0}^{n_{j}}\frac{[-2\mu_{j}^2(1-\cos\omega_{j}\tau)]^l}{(l!)^2}\frac{n_{j}!}{(n_{j}-l)!}\nonumber\\
&=& \mathrm{exp}\left[-\mu_{j}^2(1-e^{-i\omega_{j}\tau})-2n_B^j\mu_{j}^2(1-\cos\omega_{j}\tau)\right]
\end{eqnarray}
with $n_B^{j}$ the Bose distribution function $1/(e^{\beta\omega_{j}}-1)$.
We finally find
\begin{equation}
\Phi_{-+}(\tau)~=~\mathrm{exp}[-K(\tau)],
\end{equation}
where
\begin{equation}\label{eq:k}
K(\tau)~=~\sum_{j}4\frac{g_{j}^2}{\omega_{j}^2}\left[\coth\left(\frac{\omega_{j}}{2T}\right)(1-\cos\omega_{j} \tau)+i\sin\omega_{j} \tau\right].
\end{equation}
Using the spectral function $J(\omega)$ of the bath,
we can rewrite Eq. (\ref{eq:k}) as
\begin{equation}
K(\tau)~=~\frac{2}{\pi}\int_0^{\infty}d\omega\frac{J(\omega)}{\omega^2}\left[\coth\left(\frac{\omega}{2T}\right)(1-\cos\omega \tau)+i\sin\omega \tau\right].
\end{equation}
It is easy to confirm that $\Phi_{+-}(\tau)=\Phi_{-+}(\tau)$. By replacing $\tau$ with $-\tau$, we can obtain the explicit forms of $\tilde{\Phi}_{+-}(\tau)$ and $\tilde{\Phi}_{-+}(\tau)$.

Similarly, following the above procedures, we find
\begin{eqnarray}
\Phi_{++}(\tau) &=& \Phi_{--}(\tau)\nonumber\\
&=& \frac{\Delta^2}{4}\mathrm{exp}\left\{-\frac{2}{\pi}\int_0^{\infty}d\omega\frac{J(\omega)}{\omega^2}\left[-i\sin\omega \tau\right.\right.\nonumber\\
&&\left.\left.+\coth\left(\frac{\omega}{2T}\right)(1+\cos\omega \tau)\right]\right\},\nonumber\\
\tilde{\Phi}_{++}(\tau) &=& \tilde{\Phi}_{--}(\tau)\nonumber\\
&=& \frac{\Delta^2}{4}\mathrm{exp}\left\{-\frac{2}{\pi}\int_0^{\infty}d\omega\frac{J(\omega)}{\omega^2}\left[i\sin\omega \tau\right.\right.\nonumber\\
&&\left.\left.+\coth\left(\frac{\omega}{2T}\right)(1+\cos\omega \tau)\right]\right\}.
\end{eqnarray}

\section{Real part of the self-energy}\label{sec:re_se}
In this part, we will derive the real parts of the self-energies $\Sigma_{\eta}^{r/a}$. Since $\Sigma_{\eta}^{r}$ and $\Sigma_{\eta}^{a}$ are complex-conjugated, the real parts are the same and can be obtained as
\begin{equation}
\Lambda(\omega)~=~\frac{1}{2}\left[\Sigma_{\eta}^r(\omega)+\Sigma_{\eta}^a(\omega)\right].
\end{equation}
Applying Langreth theorems \cite{Haug.96.NULL} to the self-energy [c.f. Eq. (\ref{eq:se_t})], we find
\begin{eqnarray}
\Sigma_{\eta}^r(\omega)+\Sigma_{\eta}^a(\omega) &=& i\frac{\Delta^2}{4}\left[\left(_{11}G_{B,0}^{<}+_{22}G_{B,0}^{<}\right)\circ\left(G_{\eta_y,0}^r+G_{\eta_y,0}^a\right)\right.\nonumber\\
&&\left(_{11}G_{B,0}^{r}+_{22}G_{B,0}^{r}\right)\circ G_{\eta_y,0}^r-\left(_{11}G_{B,0}^{a}+_{22}G_{B,0}^{a}\right)\circ G_{\eta_y,0}^a\nonumber\\
&&+\left.\left(_{11}G_{B,0}^{r}+_{22}G_{B,0}^{r}+_{11}G_{B,0}^{a}+_{22}G_{B,0}^{a}\right)\circ G_{\eta_y,0}^<\right]
\end{eqnarray}
in the frequency domain, where $\circ$ denotes the convolution. In deriving the above result, we have used the fact that $G_{\eta_y,0}=G_{\eta_x,0}$ according to Eq. (\ref{eq:gxy}). Using relations between GFs, the above result can be further simplified as
\begin{eqnarray}
\Sigma_{\eta}^r(\omega)+\Sigma_{\eta}^a(\omega) &=& i\frac{\Delta^2}{4}\left[\sum_{\alpha=11,22}\left(_{\alpha}G_{B,0}^r+ _{\alpha}G_{B,0}^a\right)\circ\left(i\mathrm{Im}G_{\eta_y,0}^r+G_{\eta_y,0}^<\right)\right.\nonumber\\
&&+\left.\sum_{\alpha=11,22}\left(_{\alpha}G_{B,0}^> +_{\alpha}G_{B,0}^<\right)\circ\mathrm{Re}G_{\eta_y,0}^r\right],
\end{eqnarray}
where "$\mathrm{Im}$" and "$\mathrm{Re}$" denote the imaginary and real part, respectively.

Using the definition for retarded and advanced GFs of boson, we have
\begin{equation}
_{\alpha}G_{B,0}^r=\left(_{\alpha}G_{B,0}^a\right)^{\ast}=\left.\left[_{\alpha}G_{B,0}^>(\lambda)-_{\alpha}G_{B,0}^<(\lambda)\right]\right|_{\lambda=-i\omega},
\end{equation}
where we have introduced the Laplace transform
\begin{equation}
A(\lambda)~=~\int_0^{\infty}A(\tau)e^{-\lambda\tau}d\tau.
\end{equation}
Together with the results of free GFs, we find
\begin{eqnarray}
&&i\frac{\Delta^2}{4}\sum_{\alpha=11,22}\left(_{\alpha}G_{B,0}^r+ _{\alpha}G_{B,0}^a\right)\circ\left(i\mathrm{Im}G_{\eta_y,0}^r+G_{\eta_y,0}^<\right)\nonumber\\
&=& -\mathrm{Im}\left.\left[\Phi_{+-}(\lambda)-\tilde{\Phi}_{+-}(\lambda)\right]\right|_{\lambda=-i(\omega+\varepsilon)}\nonumber\\
&&+\mathrm{Im}\left.\left[\Phi_{+-}(\lambda)-\tilde{\Phi}_{+-}(\lambda)\right]\right|_{\lambda=-i(\omega-\varepsilon)},
\end{eqnarray}
and
\begin{eqnarray}
&&i\frac{\Delta^2}{4}\sum_{\alpha=11,22}\left(_{\alpha}G_{B,0}^> +_{\alpha}G_{B,0}^<\right)\circ\mathrm{Re}G_{\eta_y,0}^r\nonumber\\
&=& \frac{1}{\pi}\mathcal{P}\int\left[\Phi_{+-}+\tilde{\Phi}_{+-}\right]_{\omega-\omega^{\prime}}\frac{\omega^{\prime}}{\omega^{\prime 2}-\varepsilon^2}d\omega^{\prime}.
\end{eqnarray}
Thus we arrive the final form of $\Lambda$ [c.f. Eq. (\ref{eq:eta_ra_real}) in the main text].

\section{Evaluation of Eq. (\ref{eq:sz_kernel})}\label{sec:kernel}
Noting that $\Phi_{+-}(\tau)+\tilde{\Phi}_{+-}(\tau)$ is an even function of $\tau$, then we have $\Phi_{+-}(\omega)+\tilde{\Phi}_{+-}(\omega)=2\mathrm{Re}\left.[\Phi_{+-}(\lambda)+\tilde{\Phi}_{+-}(\lambda)]\right|_{\lambda=-i\omega}$.  Therefore, using Kramers-Kronig relations, the integral in $\Lambda$ [Eq. (\ref{eq:eta_ra_real})] can be simplified as
\begin{eqnarray}
&&\frac{1}{\pi}\mathcal{P}\int\,\left[\Phi_{+-}(\omega-\omega^{\prime})+\tilde{\Phi}_{+-}(\omega-\omega^{\prime})\right]\frac{\omega^{\prime}}{\omega^{\prime 2}-\varepsilon^2}d\omega^{\prime}\nonumber\\
&=& \mathrm{Im}\left.[\Phi_{+-}(\lambda-i\varepsilon)+\tilde{\Phi}_{+-}(\lambda-i\varepsilon)]\right|_{\lambda=-i\omega}\nonumber\\
&&+\mathrm{Im}\left.[\Phi_{+-}(\lambda+i\varepsilon)+\tilde{\Phi}_{+-}(\lambda+i\varepsilon)]\right|_{\lambda=-i\omega}.
\end{eqnarray}
Inserting it into Eq. (\ref{eq:eta_ra_real}) and using Eq. (\ref{eq:eta_ra}), we find
\begin{equation}\label{eq:sz_denom}
2i\Lambda(\omega)+\Gamma(\omega) ~=~ 2\left.\left[\tilde{\Phi}_{+-}(\lambda-i\varepsilon)+\Phi_{+-}(\lambda+i\varepsilon)\right]\right|_{\lambda=-i\omega}.
\end{equation}
Since $S_z(\omega)=2\mathrm{Re}S_z(\lambda=-i\omega)$, we deduce that $g_z(\lambda=-i\omega)=2i\Lambda(\omega)+\Gamma(\omega)$. Thus from Eq. (\ref{eq:sz_denom}) we can obtain the form of $g_z(\lambda)$ [c.f., Eq. (\ref{eq:sz_kernel})] in the main text.

\end{document}